\documentclass[iop]{emulateapj}             

\usepackage{graphicx}
\usepackage{epsfig}
\usepackage{amsmath}
\usepackage{subfigure} 
\usepackage[colorlinks,citecolor=blue]{hyperref}
\usepackage{amssymb}
\usepackage{apjfonts}
\usepackage{latexsym}
\usepackage{float}
\usepackage{epstopdf}

\newcommand{\TimeScaleRatio}{t_\mathrm{cool}/t_\mathrm{ff}}

\newcommand{\ud}{d}

\newcommand{\Msun}{M_\odot}  

\newcommand{\deriv}[2]{\frac{\ud #1}{\ud #2}}

\newcommand{\BS}{BS09}

\shorttitle{Triggering and Delivery Algorithms for AGN Feedback}
\shortauthors{G. Meece et. al.}

\begin{document}

\title{Triggering and Delivery Algorithms for AGN Feedback}

\author{Gregory R. Meece\altaffilmark{1,2}, G. Mark Voit\altaffilmark{1},  Brian W. O'Shea\altaffilmark{1,3,4,5}
         } 
\altaffiltext{1}{Department of Physics and Astronomy,
                 Michigan State University,
                 East Lansing, MI 48824}
\altaffiltext{2}{Corresponding author: meecegre@msu.edu}
\altaffiltext{3}{Department of Computational Mathematics, Science and Engineering,
                 Michigan State University,
                 East Lansing, MI 48824}
\altaffiltext{4}{National Superconducting Cyclotron Laboratory,
                 Michigan State University,
                 East Lansing, MI 48824}
\altaffiltext{5}{Lyman Briggs College, Michigan State University, East Lansing, MI 48825}

\begin{abstract}
We compare several common sub-grid implementations of AGN feedback, focusing on
the effects of different triggering mechanisms and the differences between
thermal and kinetic feedback.  Our main result is that pure thermal feedback
that is centrally injected behaves differently from feedback with even a small
kinetic component.  Specifically, pure thermal feedback results in excessive
condensation and smothering of the AGN by cold gas because the feedback energy
does not propagate to large enough radii.  We do not see large differences
between implementations of different triggering mechanisms, as long as the
spatial resolution is sufficiently high, probably because all of the
implementations tested here trigger strong AGN feedback under similar
conditions.  In order to assess the role of resolution, we vary the size of the
``accretion zone'' in which properties are measured to determine the AGN
accretion rate and resulting feedback power.  We find that a larger accretion
zone results in steadier jets but can also allow too much cold-gas condensation
in simulations with a Bondi-like triggering algorithm.  We also vary the
opening angle of jet precession and find that a larger precession angle causes
more of the jet energy to thermalize closer to the AGN, thereby producing
results similar to pure thermal feedback.  Our simulations confirm that AGN can
regulate the thermal state of cool-core galaxy clusters and maintain the core
in a state that is marginally susceptable to thermal instability and
precipitation.\par
\end{abstract}

\section{Introduction}

An active galactic nucleus (AGN) is thought to be present in the core of nearly
every massive galaxy and galaxy cluster.  Based on estimates of jet power from
AGN-inflated cavities, it has become clear that an AGN can strongly influence
cooling and condensation of gas in its host galaxy
\citep[e.g.,][]{2012NJPh...14e5023M}, potentially explaining the relationships
observed between the mass of a galaxy's central black hole and the velocity
dispersion of its stars \citep[e.g.,][]{2000ASPC..197..221M,
2000ApJ...539L...9F}, as well as the star-formation properties of galaxies with
AGNs \citep[e.g.,][]{2003MNRAS.346.1055K}.  Directly simulating the
co-evolution of AGNs together with their host galaxies is not computationally
feasible due to the large differences in mass and size between an AGN and its
host galaxy.  Also, the diverse set of complex physical processes that govern
AGN accretion and outflow production remain poorly understood.  To circumvent
these difficulties in studies of galaxy evolution, simulators have developed a
number of ``subgrid" implementations of AGN feedback that are intended to
capture the interplay between the AGN and its environment without representing
the details of AGN accretion on smaller scales \citep[see, for
example][]{2004MNRAS.348.1105O, 2005MNRAS.361..776S, 2008ApJ...687L..53P, 2009MNRAS.398...53B,
2011MNRAS.411..349G, 2012MNRAS.420.2662D, 2014ApJ...789...54L, 2015MNRAS.448.1504S}.  However,
subgrid implementations of AGN feedback vary widely, and there has been little
systematic comparison \citep[but see][]{2013MNRAS.431.2513W,
2012MNRAS.427.1614Y}. From the explorations of parameter space carried out in
these studies, it has become evident that varying certain AGN feedback
parameters can lead to strong differences in feedback power and energy
propogation. In this paper, we compare several of the popular methods for
implementing AGN feedback.\par

An AGN consists of a supermassive black hole (SMBH) surrounded by a disk of
accreting material. Twisted magnetic fields in the disk are thought to channel
charged particles into jets, which can draw additional power from the spin of
the SMBH via the Blandford-Znajek effect \citep{1977MNRAS.179..433B,
1982MNRAS.199..883B}. These relativistic jets produce synchrotron emission,
which is observed in the radio band.  Hence, this mode of AGN energy output is
termed ``radio-mode'' feedback \citep[e.g.,][]{2001ApJ...554..261C,
2005Natur.435..629S}.  Additionally, differential rotation in the accretion
disk will heat the accreting material, producing a strong UV flux. If the SMBH
is accreting near the Eddington Limit, radiation pressure may also drive large
outflows. This ``quasar-mode'' feedback \citep[e.g.,][]{2005MNRAS.363L..91C} is
composed of non-relativistic material and is more isotropic than radio-mode
feedback.\par

The length and mass scales that are important for AGN are much smaller than the
range covered by galaxies and galaxy clusters. For example, a black hole with a
mass similar to the one dwelling at the center of the Perseus Cluster
\citep[$\sim 3.4 \times 10^8 \, M_\odot$; see][]{2005MNRAS.359..755W} has a
Schwarzchild radius of only a few AU, whereas the virial radius of a galaxy
cluster is $\gtrsim 1$~Mpc.  Cosmological simulations capable of modeling
entire galaxy clusters typically have a maximum resolution of $\sim 1$~kpc,
meaning that the AGN's behavior must be approximated with a subgrid model.
Furthermore, AGN are known to be variable on timescales much shorter than the
dynamical time of a galaxy.  Cosmological simulations that model structure
formation over a Hubble Time must therefore rely on an AGN model that smooths
out this short-term variability while preserving the large-scale behavior of
the resulting feedback.\par

In order for a self-regulated feedback loop to arise, a subgrid AGN model must
capture the coupling between an AGN and its fuel supply.  A triggering
algorithm must somehow estimate the mass accretion rate onto the SMBH, which
translates into a proportional release of feedback energy.  Then a delivery
algorithm must prescribe how that feedback energy interacts with the local
environment.  In cosmological simulations, the subgrid AGN model must also
include prescriptions for following the creation, advection, and merger of
SMBHs, although these are not discussed in this work \citep[see][for more
discussion on these topics]{ 2007MNRAS.380..877S, 2008ApJ...676...33D,
2013MNRAS.431.2513W, 2013MNRAS.436.3031V}. Instead, we are focusing on just the
triggering and delivery algorithms.\par

There are a number of reasons to believe that AGN feedback in massive galaxies
is self-regulated.  First, as stated earlier, strong relationships have been
found between the masses of SMBHs and the properties of their host galaxies.
Second, AGNs are considered the best candidates for solving the ``Cooling Flow"
problem in galaxy clusters and elliptical galaxies \citep{1995MNRAS.276..663B,
2007ARA&A..45..117M, 2013AN....334..386N}.  Many galaxy clusters have central
cooling times that are far shorter than the age of the clusters, but the
observed star-formation rates are an order of magnitude or more below what would be
expected from uninhibited cooling \citep{2008ApJ...681.1035O,
2010ApJ...719.1619O, 2011ApJ...734...95M}.  Also, the amounts of cold gas that
seem to be accumulating are much less than one would naively expect
\citep{2003ApJ...590..207P,2006PhR...427....1P}.  This tension implies the
existence of a heat source that roughly balances cooling losses. Non-AGN heat
sources, such as supernovae, mergers, conduction, and preheating have been
proposed, but are either not powerful enough to balance cooling
\citep[e.g.,][]{2013ApJ...763...38S} or are inconsistent with observations.
AGNs, however, are known to be present in the central galaxies of galaxy
clusters and produce feedback energy comparable to the cooling rate.
\citet{1995MNRAS.276..663B} showed that when cooling-triggered jets are added
to models of cool-core clusters, alternating periods cooling and jet heating
can lead to a quasi-steady state for the ICM in the cluster core.  As discussed
in \citet{2012NJPh...14e5023M}, the jet power must be closely coupled to the
cooling rate if AGN are balancing cooling in clusters.  Otherwise, the AGN
would either overheat or underheat the intra-cluster medium (ICM).\par

One simple algorithm for estimating the AGN accretion rate bases the scaling
properties on the Bondi-Hoyle accretion model, set out in
\citet{1952MNRAS.112..195B}, which implicitly assumes that the SMBH is
accreting hot ambient gas directly from the ICM.  Strictly speaking, such a
Bondi accretion flow should be steady-state, spherically symmetric, and
isentropic.  Accretion then becomes supersonic within the Bondi radius given by
$R_{\rm Bondi} \approx {2GM_{\rm BH}}/{c_{s}^2}$ where $M_{\rm BH}$ is the
black hole mass and $c_{s}$ is the sound speed of the gas near $R_{\rm Bondi}$,
and proceeds at a rate $\dot{M}_{\rm Bondi}$ that depends on $M_{\rm BH}$ and
$c_s$.  However, the Bondi radius is unresolved in many numerical simulations
of AGN feedback, as is its impact on galaxy evolution.
\citet{2005MNRAS.361..776S} therefore proposed a parameterized Bondi accretion
rate with an artificial boost factor $\alpha$ such that $\dot{M}_{\rm BH} =
\alpha \dot{M}_{\rm Bondi}$.  The boost factor $\alpha$ is typically chosen to
be large because the actual gas properties at the Bondi radius are likely to
permit a greater accretion rate than would arise in under-resolved simulations.
\citet{2005MNRAS.361..776S} and subsequent studies following up on that work
use $\alpha=100$, while \citet{2008MNRAS.387...13K} use $\alpha=300$.
\citet{2006ApJS..166....1H}, in contrast, use a factor that is near unity,
albeit for studies of galaxy-scale phenomena.\par

In reality, the assumptions of Bondi accretion --- steady homogeneous flow,
spherical symmetry, and adiabaticity --- are unlikely to be valid near the
Bondi radius around a massive galaxy's SMBH \citep[see, for example, the
discussion in][]{2012ApJ...754..154M}.  Furthermore, models relying on
standard Bondi accretion have problems generating sufficiently powerful
outflows without a large boosting factor (see \BS\ for discussion) and with
reproducing the properties of observed cool-core clusters absent fine tuning.
An alternative model for self-regulated accretion, described by
\citet{2005ApJ...632..821P}, posits that the AGN is primarily fueled by
accretion of cold, dense gas that rains down in a stochastic manner. Aside from
the prior assumption that the AGN heating rate is linked to cooling in the ICM,
this model is supported by observations of cold gas and star formation which
indicate that at least some gas is able to cool
\citep[e.g.,][]{2001MNRAS.328..762E, 2008ApJ...683L.107C, 2008ApJ...681.1035O}.
In this ``moderate cooling flow'' or ``cold feedback'' model, radial mixing
resulting from strong AGN outbursts creates large inhomogeneities that cool and
condense at radii between 5 and 30 kpc from the AGN.  The condensates then rain
down on the SMBH, powering subsequent outbursts.  This model couples the AGN to
the cooling properties of the entire cluster core rather than only to the
region directly surrounding the SMBH.  Importantly, the coupling also occurs
over timescales longer than the freefall in the core, leaving time for gas to
cool and condense. Cold mode feedback has been implemented in recent
simulations, notably those of \citet{2011MNRAS.415.1549G,2011MNRAS.411..349G,
2012ApJ...746...94G, 2014ApJ...789...54L, 2014ApJ...789..153L,
2015ApJ...811...73L}, which attain self-regulated states similar to those
observed in galaxy-cluster cores.\par

To mimic the effects of both cold and hot accretion modes, \citet[hereafter
referred to as \BS]{2009MNRAS.398...53B} proposed a model that invokes Bondi
accretion with a density-dependent boost factor. This boost factor equals unity
at low densities, giving the classical Bondi accretion rate, but ramps up
quickly above a pre-chosen density threshold in order to account, rather
crudely, for cooling, condensation, and accretion of condensed gas.\par

In the simplest models for delivery of AGN feedback, all of the feedback energy
is assumed to thermalize at scales below the resolution of the grid and is
deposited as thermal energy in a small central region. This approach is used in
\citet{2005MNRAS.361..776S} and subsequent works, including recent simulations
such as the Illustris simulation \citep{2014MNRAS.444.1518V}, Rhapsody-G
\citep{2015arXiv150904289H} and the simulations of \citet{2015ApJ...813L..17R}.
In reality, AGN outflows are likely asymmetric on scales of several kpc with a
significant proportion of their energy in kinetic form.  Such bipolar outflows
may be important for transporting feedback energy to large distances from the
AGN and for mixing metals out to distances of $\sim 100$~kpc from the central
galaxy \citep{2011ApJ...731L..23K, 2015MNRAS.452.4361K}.  Although much work has been done studying
highly collimated outflows on small scales over short time periods
\citep{2006ApJ...645...83V, 2004MNRAS.348.1105O}, it is not straightforward to
implement them in large-scale simulations with coarser resolution.\par

Given the increasing awareness that a proper treatment of AGN feedback is
essential for accurate modeling of the evolution of large galaxies, it is
important that the consequences of different AGN feedback implementations be
understood.  The rest of this paper compares several commonly used algorithms
for triggering and delivery of AGN feedback in the context of an idealized
galaxy-cluster core, in order to explore how they differ in representing the
coupling of an AGN to its environment, the total AGN feedback energy produced,
and the resulting thermodynamic profiles of the ambient medium.  Section
\ref{section:method} discusses our simulation setup and outlines the triggering
and delivery methods we study.  Section \ref{section:results} describes the
results of changing the triggering and delivery algorithms. Section
\ref{section:discussion} discusses our results in the context of
thermal-instability analyses of cold gas accumulation and self regulation,
along with a discussion of how physical processes that were not included might
have affected our results if they had been included.  Finally, Section
\ref{section:conclusions} summarizes the key results and points out avenues and
opportunities for further study.

\section{Method}\label{section:method}

In this work, we consider the interplay between ICM cooling and AGN feedback
using a simplified AGN model in an idealized galaxy cluster environment. The
simulations are performed using the adaptive mesh hydrodynamics code
\texttt{Enzo}\footnote{http://enzo-project.org/} \citep{2014ApJS..211...19B}
and analyzed using the \texttt{yt}\footnote{http://yt-project.org/} analysis
toolkit \citep{2011ApJS..192....9T}.\par

\subsection{Simulation Environment}
Our simulations include hydrodynamics, gravity, radiative cooling, and AGN
feedback. We use a static gravitational potential representing both the cluster
and its BCG but do not account for the self-gravity of the gas, which we assume
to be negligible.  We use a tabulated cooling function taken from
\citet{2009A&A...508..751S}, assuming a uniform metallicity of half the Solar
value.  This cooling function does not allow gas to cool below $10^4$ K, which
does not affect the qualitative behavior of our simulations, since any
processes occurring at lower temperatures would take place below our spatial
resolution limit. For analysis purposes, we define any gas below $3 \times
10^4$ K as ``cold.'' Section \ref{section:additional_physics} discusses the
potential effects of including additional physical processes such as magnetic
fields, conduction, and star formation, which may affect AGN feedback but are
not included in our simulations.\par

Unless otherwise noted, the simulation setup encompassed a box of length 3.2
Mpc per side with a $64^3$ cell root grid and 8 levels of AMR refinement,
giving a maximum spatial resolution of 196~pc.  A set of 8 nested grids, centered on the
cluster core, with twice the resolution and half the width of the previous
level, were created during initialization and were never de-refined. Additional
refinement was allowed to occur based on strong density or energy gradients,
baryon overdensity, and cooling. All cells containing material that was ejected
from the central 10~kpc, as indicated by a passive tracer field added to
material within that region, were covered by at least 4 levels of refinement.
Finally, the zone around the AGN where accretion was measured and feedback was
applied was always refined to the maximum level.  We do not take cosmological
expansion into account.\par

\subsection{Cluster Setup}
Following the work of \citet{2012ApJ...747...26L}, we initialize the ICM as a
hydrostatic sphere of gas within a static spherical gravitational potential.
The gravitational potential comprises two components: an NFW halo and the
stellar mass profile of the BCG.  The virial mass $M_{200}$ and concentration
parameter $c$ of the NFW halo are defined with respect to the radius within
which the mean mass density is 200 times the critical density.  For the BCG we
assume a mass profile of the form
\begin{equation}
   M_*(r) = M_{4} \left[\frac{2^{-\beta_*}}{\left(r/4\, \rm{ kpc}\right)^{-\alpha_*} \left(1 +
   r/4\, \rm{ kpc}\right)^{\beta_*-\alpha_*}}\right]\,\,\,,
\end{equation}
where $M_4$ is the stellar mass within 4 kpc and $\alpha_*$ and $\beta_*$ are
constants. As in \citet{2012ApJ...747...26L} and \citet{2006ApJ...638..659M},
we used the Perseus cluster as a template, choosing $M_{200}=8.5 \times 10^{14}
\Msun$, $c=6.81$ for the NFW halo, $M_4=7.5 \times 10^{10} \Msun$,
$\alpha_*=0.1$, and $\beta_*=1.43$ for the BCG. With these mass profiles, the
BCG is gravitationally dominant $\lesssim10$~kpc from the center, while outside
of this radius the NFW halo dominates the potential.\par

Although we do not take cosmological expansion into account in our simulations,
we do use a vanilla $\Lambda$CDM model in order to specify the virial mass
of the NFW halo and to set its gas temperature.
For initialization, we assume a cluster at redshift $z=0$ and a cosmology with
$\Omega_M=0.3$, $\Omega_\Lambda=0.7$, and $H_0=70$ km/s/Mpc. We do not expect our
results to be sensitive to small changes in these parameter values.\par

The hydrostatic gas in the halo is initialized with an entropy
profile of the form
\begin{equation}\label{eqn:entropy_profile}
   K(r) = K_0 + K_{100} (r/\rm{100 kpc})^{\alpha_K}
\end{equation}
where we use the definition of specific entropy used in the ACCEPT database \citep{2009ApJS..182...12C}:
\begin{equation}\label{eqn:entropy}
   K \equiv \frac{k_B T}{n_e^{2/3}} \,\,\,.
\end{equation}
For the Perseus cluster, ACCEPT gives values of $K_0=19.38$ keV cm$^2$, $K_{100}=119.87$ keV
cm$^2$, and $\alpha_K=1.74$, and we use them for our initial configuration. 
The condition for hydrostatic equilibrium is
\begin{equation}\label{eqn:HSE}
   \deriv{P}{r} = -\rho g \,\,\,.
\end{equation}
Together, the the specified entropy profile and the hydrostatic condition (equations
\ref{eqn:entropy_profile}, \ref{eqn:entropy}, and \ref{eqn:HSE}) give a
differential equation relating temperature and entropy. It remains to specify a
boundary condition so that this equation can be integrated. Following
\citet{2005RvMP...77..207V}, the temperature of a hydrostatic ICM can be
approximated as
\begin{equation}
   k_B T_{200} = \frac{\mu m_p}{2} \left[10 G M_{200} H(z)\right]^{2/3}
\end{equation}
We take this as a characteristic temperature for the ICM near the virial radius
and integrate inwards and outwards to find the temperature and density profiles
for the rest of the cluster.\par

\subsubsection{Tracer Fluid}
In order to track the gas directly affected by AGN feedback, we continuously
inject a (passive) tracer fluid into the central 10~kpc.  This passive tracer
also allows us to measure the radial extent of feedback heating and also
indicates the amount of metal transport facilitated by AGN jets and rising
cavities, which are thought to play an important role in shaping the
metallicity profiles of clusters. The amount of tracer injected per unit mass
$\Delta \rho_T$ is given by
\begin{equation}
   \Delta \rho_{T} = \text{SSFR} * Y
\end{equation}
where we assume a specific star-formation rate SSFR~$=10^{-11}$ yr$^{-1}$ and a
yield $Y=0.02$.  These assumptions are meant to be a crude approximation for
metal injection by the old stellar population of the BCG.  We emphasize that all we are doing is
injecting passive tracer fluid.  No actual star formation takes place, and the tracer fluid 
does not affect the radiative cooling rate.  Our primary interest is 
radial transport and distribution of the tracer fluid.   We do not expect 
its concentration to match metallicity values in the ICM of observed clusters.

\subsection{Feedback and Jet Modeling}
AGN are complicated systems governed by physical processes that are poorly
constrained and span many orders of magnitude in space and time. Our goal here
is not to understand all the details of AGN physics but rather to study the
interplay between accretion, jet outflows, and the thermal state of the ICM. To
this end, we implement a simplified ``AGN Particle'' model, wherein accretion
onto the AGN launches outflows that are insensitive to the details of gas
accretion on scales $<200$~pc.  We implement several triggering mechanisms,
each with a different algorithm for determining the accretion rate $\dot{M}$
into the region surrounding the central supermassive black hole, which sets the
scale of the AGN feedback response.  In each case, the resulting output of
feedback power is taken to be $\dot{E}=\epsilon \dot{M} c^2$, where $\epsilon$
is a feedback efficiency factor and $c$ is the speed of light.  The accretion
rate $\dot{M}$ is not necessarily the actual accretion rate onto the central
black hole, and in our idealized implementations no gas is removed from the
simulation volume.  Instead, it is assumed to be reheated and expelled from the
vicinity of the black hole by feedback.  Regardless of the triggering
mechanism, precessing jets are launched from disk-shaped regions on either side
of the AGN as described in the following subsections.  Please refer to Table
\ref{table:parameter_table} for fiducial values of the AGN feedback
parameters.\par

\subsubsection{Triggering Mechanisms}
Each of the following triggering methods calculates $\dot{M}$ and removes gas mass from the
grid within a specified radius given by the parameter $R_{\rm acc}$.

\paragraph {\it Cold-Gas Triggered Feedback} Meant to replicate the triggering mechanism
used in \citet{2014ApJ...789...54L}, feedback is triggered by the presence of
gas within $R_{\rm acc}$ and at or below a threshold temperature $T_{\rm{floor}}$. 
The accretion rate corresponding to a single cell is
\begin{equation}
   \dot{M}_{\rm{cell}} = \frac{M_{\rm{cell}}}{t_{\rm{acc}}}
\end{equation}
where $t_{\rm{acc}}$ is a constant timescale. Following
\citet{2014ApJ...789...54L}, we choose $t_{acc}=5 \rm{ Myr}$, which is close to
the average freefall time near the accretion radius.\par

\paragraph {\it Boosted Bondi-like Triggering} The accretion rate is set to the
Bondi accretion rate derived from conditions within $R_{\rm acc}$ 
and multiplied by a constant boost factor $\alpha$ so that
\begin{equation}
   \dot{M} = \alpha \frac{2 \pi G^2 M_{\rm BH}^2 \hat{\rho}}{(\hat{v}^2 +
   \hat{c_s}^2)^{3/2}}
\end{equation}
where $G$ is the gravitational constant, $M_{\rm BH}$ is the mass of the black
hole, and $\hat{\rho}$, $\hat{v}$, and $\hat{c_s}$ are the mass- averaged
density, velocity magnitude, and sound speed within $R_{\rm acc}$. In this work
we adopt $\alpha=100$. Mass is removed from each cell within $R_{\rm acc}$ in a
mass-averaged sense, such that
\begin{equation}
   \Delta M_{\rm{cell}} = \frac{M_{\rm{cell}}}{M(<R_{\rm acc})} \dot{M} \, \Delta t
\end{equation}

\paragraph {\it Booth and Schaye Accretion} As described in
\BS, the accretion rate follows the Bondi formula but
with a boost that depends on the gas density as
\begin{equation}\label{eqn:BS_formula}
 \alpha = \begin{cases} 
   1 & n \leq n_0\\
   (n/n_0)^\beta & n > n_0\\
\end{cases}
\end{equation}
Following \cite{2009MNRAS.398...53B}, we take $n_0=0.1$ cm$^{-3}$ and $\beta=2$.

\begin{deluxetable}{ccc}
 \tablecaption{Fiducial AGN Feedback Parameters}
     \tablehead{\colhead{Parameter} & \colhead{Value} & \colhead{Description} }
     \startdata
      $\epsilon$ & $10^{-3}$ & Jet efficiency\\
      $R_{\rm acc}$ & 0.5 kpc & Accretion radius\\
      $T_{\rm{Floor}}$ & $3 \times 10^4$ K & Temperature floor\\
      $M_{\rm BH}$ & $1.0 \times 10^{8} \Msun$ & SMBH mass\\
      $\phi_{\rm{Jet}}$ & 0.15 radians & Jet precession angle \\
      $\tau_{\rm{Jet}}$ & 10 myr & Jet precession period\\
      $R_J$ & 0.5 kpc & Radius at which jets are launched\\
      $R_D$ & 0.5 kpc & Initial radial thickness of jets
    \enddata
 \tablecomments{These parameter values are used for all simulations unless otherwise noted in the text. \vspace*{1em}}
   \label{table:parameter_table}
\end{deluxetable}

\subsubsection{Jet Implementation}
After the total accretion rate $\dot{M}$ during a timestep $\Delta t$ is 
calculated with one of these triggering methods, a corresponding amount of
feedback energy $\epsilon \dot{M} c^2 \,\Delta t$ is added to the ejected gas.
We assume the ejected mass to be equal to $\dot{M} \Delta t$, which is an
idealization. In reality, the mass-loading factor of the jets will depend on subgrid physics 
that is not yet well understood.  However, \citet{2012MNRAS.420.2662D} find that 
the choice of mass-loading factor does not strongly affect their results.

A fraction $f_{\rm{k}}$ of the feedback energy is added to the ejected mass as kinetic energy, 
while the rest is added as thermal energy. This naturally results in a jet velocity of
\begin{equation}
   v_{\rm{Jet}} = c \sqrt{2 \epsilon f_{\rm{kinetic}}}
\end{equation}
or around $v \approx 0.045 c$ for our fiducial parameter choices.  Kinetic
energy and the associated mass are put into the grid through two disks each of
radius $R_D$ located on either side of the AGN at a distance $R_J$ from the
center.  The jets are oriented at a fixed angle $\phi_{\rm{Jet}}$ with respect
to the $z$ axis and precess around it with a period $\tau_{\rm{jet}}$. \par

For simulations with pure thermal feedback ($f_{\rm k} = 0$), we again follow
the method of \BS\ in order to prevent the injected thermal energy from
immediately being radiated away.  Feedback energy is stored up until enough
accumulates to heat the gas in the injection zone to at least $T_{min}=10^7$ K.
Exploratory simulations with $T_{min}=10^8$ did not show a noticeable difference
in behavior, in agreement with \BS.  We observe that this algorithm results in a
series of thermal pulses as AGN feedback is ramping up, but comes close to
steady injection when the AGN power is high. We performed tests using this
injection threshold with some kinetic feedback ($f_K > 0.0$) but did not observe
a noticeable difference when compared to simulations with continuous energy injection.\par

Unless otherwise noted we use the parameters given in Table
\ref{table:parameter_table} for all simulations.\par

\newpage
\subsection{Hydro Method}
The simulations in this work use a 3D version of the ZEUS hydrodynamics method
\citep{1992ApJS...80..753S} because of its robustness and speed. ZEUS is known
to be a relatively diffusive method and requires an artificial viscosity term
that may affect the accuracy of our hydrodynamics calculations. We have
experimented with using a piecewise-parabolic method (PPM)
\citep{1984JCoPh..54..174C}, but encountered numerical difficulties relating to
the strong discontinuities occurring at the injection site.\par

\section{Results}\label{section:results}

The most striking differences within our suite of simulations  
are between AGN feedback algorithms that deliver all of the feedback in thermal form ($f_{\rm k} = 0$)
and those that deliver at least some kinetic feedback ($f_{\rm k} > 0$).
Changes in the triggering method produce smaller differences in qualitative behavior,
probably because all three triggering methods implemented here end up strongly 
boosting the feedback response when significant amounts of cold gas accumulate
near the central black hole. We will therefore present our results on delivery mechanisms 
first and triggering mechanisms second.

\subsection{Delivery of Feedback:  Thermal vs. Kinetic}

Injection of AGN feedback energy heats the surrounding gas through several processes.
First, if $f_{\rm k}<1.0$, then the AGN directly injects thermal energy into the ICM.
Second, interactions between the AGN outflow and the ICM produce shocks that propagate outward 
and heat the ambient gas in a quasi-isotropic manner. Third, outflows drive turbulence that can
heat the ICM as the turbulence decays. Finally, momentum from the AGN outflow---either
directly injected in the form of a kinetic jet or driven by thermal expansion of hot 
bubbles---can dredge low-entropy gas out of the core and mix it with higher entropy gas 
at larger radii.\par

\begin{figure}
   \includegraphics[width=0.49\textwidth]{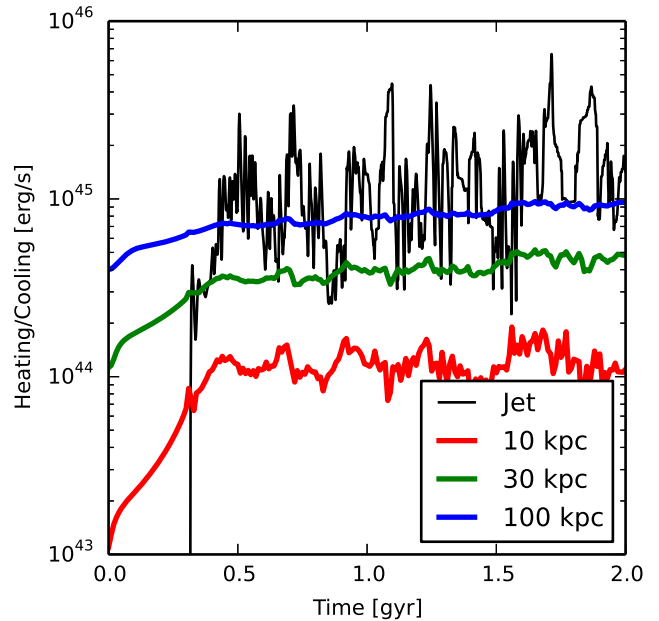}
   \centering
   \caption{Total AGN power (thermal + kinetic)  and cooling luminosity for a
simulation with cold gas triggered feedback and $f_{\rm{k}}=0.5$. The jagged
black line shows instantaneous jet power sampled every 5 Myr. Red, green, and
blue lines show the total cooling luminosity of gas within 10, 30, and 100 kpc
respectively, sampled at the same cadence.}
   \label{fig:edot_vs_cooling}
\end{figure}

\subsubsection{Feedback Power}
All of our simulations with $f_{\rm k} > 0.0$ follow similar patterns of
evolution.  Initially, the cluster core is smooth, spherically symmetric, and
contains no cold gas.  The core gas then cools, contracts, and grows denser for 
$\sim 0.3$~Gyr until cold clouds begin to condense at the center and strongly boost the 
jet power.  Figure~\ref{fig:edot_vs_cooling} shows both the jet power and
cooling luminosity within different radii during the first 2~Gyr of a cold-gas triggered 
feedback simulation that delivers 50\% of the feedback power as kinetic energy.  
Notice that the core achieves approximate long-term balance when the jet power rises to
match the cooling luminosity from within the central $\sim 100$~kpc.  This
is typical of our simulations that have a significant fraction of the feedback power 
in kinetic form. However, the total feedback power becomes much greater in
simulations with purely thermal feedback. \par

\begin{figure*}
   \includegraphics[width=0.95\textwidth]{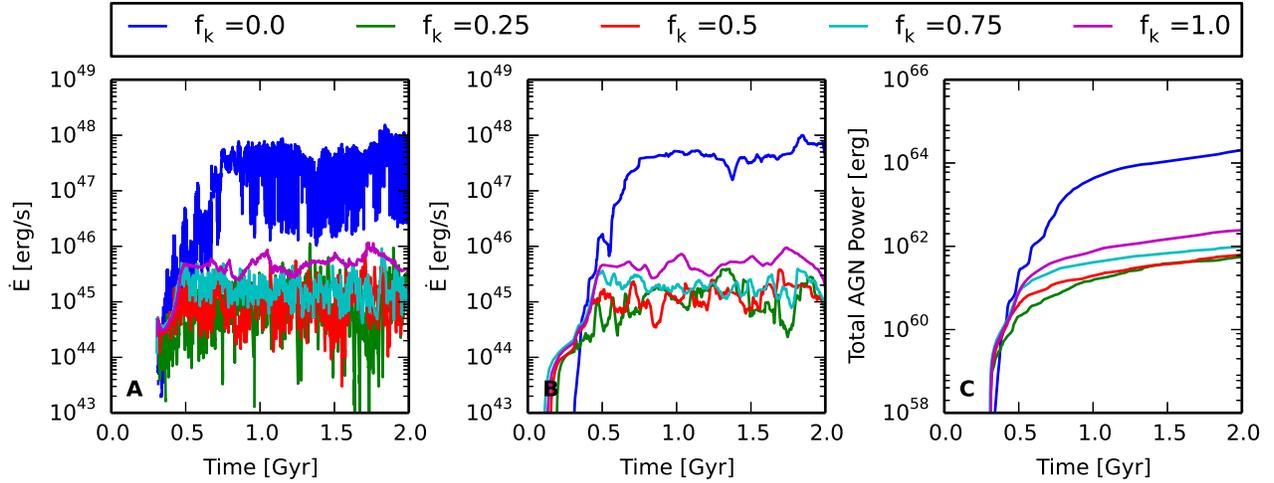}
   \centering
   \caption{Feedback power ($\dot{E}$) (thermal + kinetic) as a function of
time for simulations with cold gas triggering and varying values of $f_{\rm
k}$, the fraction of feedback power in kinetic form.  Panel A shows the
instantaneous value of $\dot{E}$. In Panel B, $\dot{E}$ has been smoothed over
a 50 Myr uniform smoothing kernel. Panel C shows the cumulative energy released
by the AGN.}
   \label{fig:fkinetic_edot}
\end{figure*}

Figure \ref{fig:fkinetic_edot} illustrates the vast difference in feedback 
power between our simulations with $f_{\rm k} = 0$ and those with $f_{\rm k} > 0$. 
Pure thermal feedback eventually saturates at a power level more than 
two orders of magnitude greater than in the simulations with some kinetic feedback,
even when compared to the case with $f_{\rm k}=0.25$.  Furthermore, it
can be seen that
the average feedback power in the self-regulated systems with at least some kinetic power
is not monotonically dependent on $f_{\rm k}$.  As long as some of the feedback
power is kinetic, self-regulation happens at a power level of $\sim 10^{45} \,{\rm erg \, s^{-1}}$,
which is similar to the time-averaged AGN power inferred from observations of X-ray cavities 
in galaxy cluster cores \citep[e.g.,][]{2012NJPh...14e5023M}.\par

\subsubsection{Cold Gas Accumulation} 

\begin{figure}
   \includegraphics[width=0.49\textwidth]{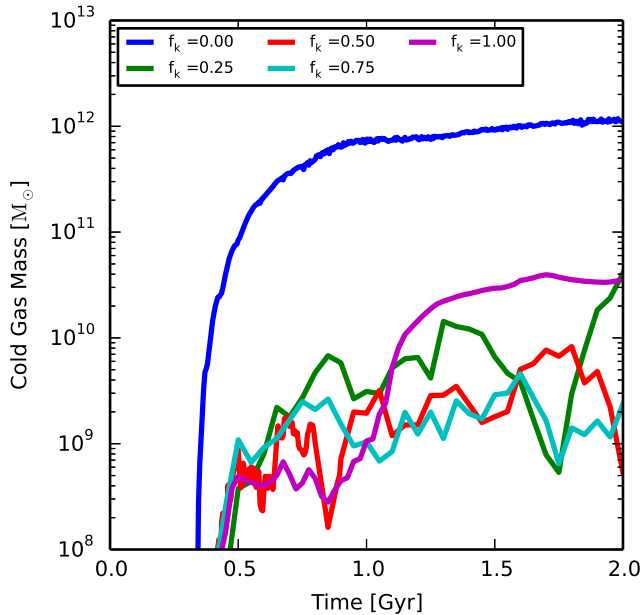}
   \centering
   \caption{
   Total mass of cold gas as function of time for simulations with 
   cold gas triggering and differing values of $f_{\rm k}$.  The amount of cold gas that
   accumulates in the simulation with $f_{\rm k} = 0$ is two orders of magnitude 
   greater than in any of the simulations with some of the feedback energy in kinetic 
   form.}
   \label{fig:cold_mass_comparison}
\end{figure}

Feedback power becomes excessively large in the $f_{\rm k} = 0$ case because
pure thermal feedback is ineffective at preventing large amounts of cold gas
from accumulating.  Figure~\ref{fig:cold_mass_comparison} shows that $\sim
10^{12} \, \Msun$ of cold gas accumulates in less than 1~Gyr when $f_{\rm k} =
0$, whereas $\lesssim 10^{10} \,\Msun$ accumulates during the same time period
in simulations with at least some kinetic power.  The large cold-gas reservoir
in the $f_{\rm k} = 0$ case is not sufficiently disrupted by thermal feedback
and therefore provides enough cold fuel for the AGN to maintain a feedback
power exceeding $10^{47} \, {\rm erg \, s^{-1}}$.  Even at this power level,
the AGN fails to eject or eliminate much of the cold gas because the cold gas
is very efficient at radiating away feedback energy owing to the $n^2$
dependence of the cooling rate. The result is that the feedback energy that
does go into the cold gas is almost immediately radiated away.  Further,
feedback energy tends to propagate more readily through the hot ambient medium
along the paths of least resistance, and ends up increasing the thermal energy
of the diffuse, volume-filling gas without diminishing the mass of cold gas
embedded within it.\par

If star formation had been allowed to proceed in our simulations, much of the cold gas that 
accumulates near the center would eventually have formed stars.  The results shown 
in Figure~\ref{fig:cold_mass_comparison} therefore indicate that that pure thermal feedback
would permit a time-averaged star-formation rate $\sim 10^{2-3} \Msun \, {\rm yr}^{-1}$
during the first $\sim 1$~Gyr, which is much larger than observed in all but the most 
actively star-forming galaxy cluster cores \citep{2008ApJ...681.1035O}.  
In order to understand why kinetic feedback is so much more successful
than thermal feedback in suppressing cold gas accumulation and the star formation
that would result, we need to look at how the choice of $f_{\rm k}$ affects the radial
distribution of density, temperature, and entropy in the hot ambient medium.\par

\subsubsection{Radial Profiles}

\begin{figure*}
   \includegraphics[width=\textwidth]{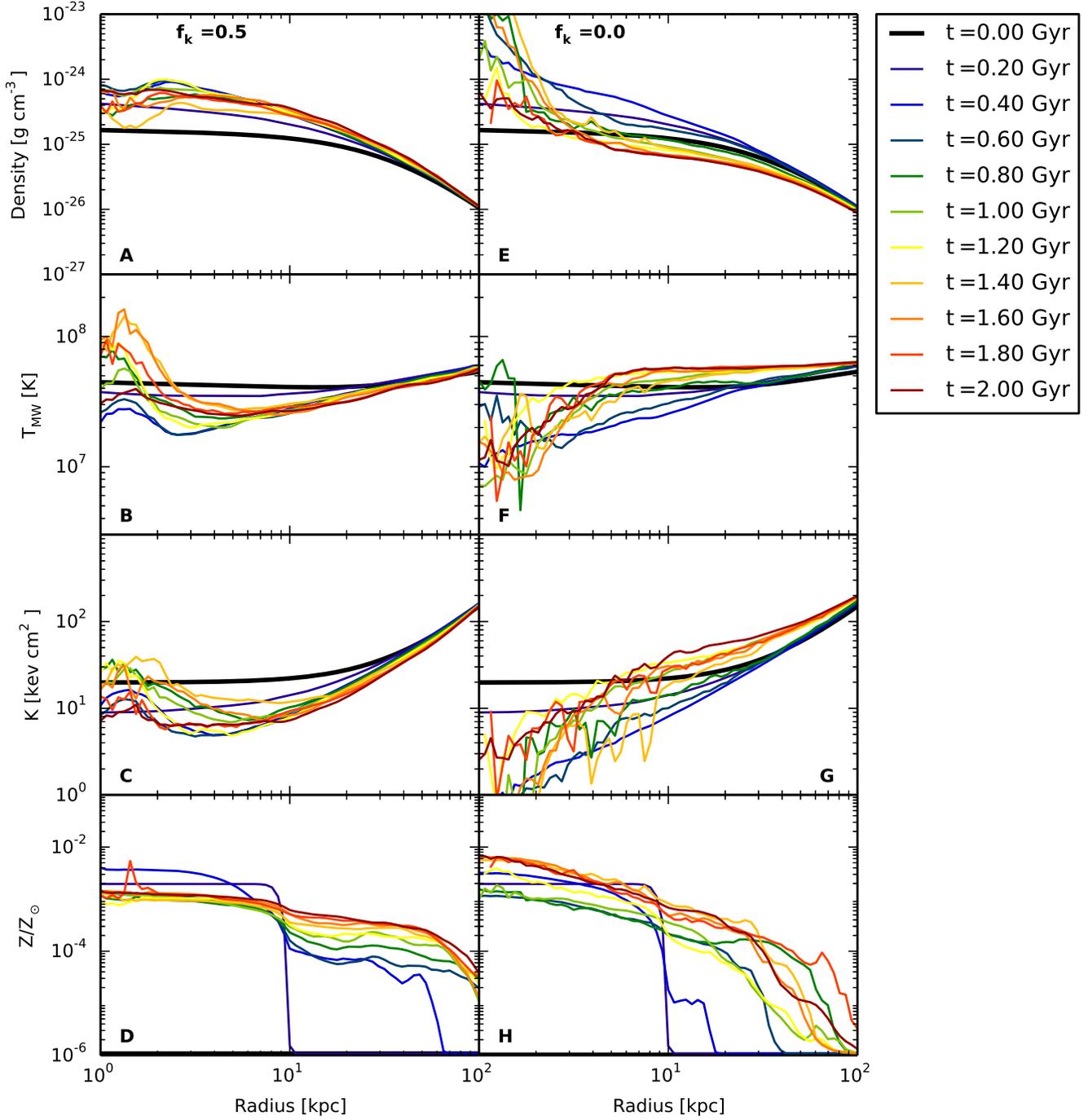}
   \centering
   \caption{Profiles of various quantities as the simulation evolves for
   simulations with $f_{\rm k}=0.5$ (left) and $f_{\rm k}=0.0$ (right) with cold gas
   triggering. Thick black lines denote the initial conditions, while other line
   colors indicate values at later times. Density is weighted by volume,
   temperature by mass, and metallicity my mass. Entropy is computed using the
   volume weighted temperature and the mass weighted density, as discussed in
   Section 3.2.3 of \citet{2013ApJ...763...38S}.}
   \label{fig:profiles_vs_time}
\end{figure*}

Figure \ref{fig:profiles_vs_time} shows how the average values of density,
temperature, entropy, and concentration of tracer fluid change over time at
each radius in simulations with mixed kinetic and thermal feedback ($f_{\rm
k}=0.5$, left panels) and pure thermal feedback ($f_{\rm k}=0$, right panels).
Gas outside of $\sim 100$~kpc is not shown because it does not evolve
appreciably over 2 Gyr, as the cooling time at large radii is long and little
of the AGN feedback energy propagates to those radii.  Inside of 100 kpc, the
$f_{\rm k}=0.5$ simulation reaches a nearly steady state in $\sim 0.5$~Gyr,
with density, temperature, and entropy continuing to fluctuate within narrow
ranges after that time.  The profiles of the tracer fluid concentration do not
reach a steady state, as the tracer is continuously injected over time and
distributed outward by the jet. However, those profiles do show that tracer
fluid is quickly mixed with the ambient gas out to $\gtrsim 50$~kpc from the
center.\par

In contrast, the simulation with $f_{\rm k} = 0$ does not reach a steady state.
In particular, the azimuthally averaged specific entropy of gas outside of the
central few kpc steadily rises with time, causing a steady drop in ambient
density and a steady rise in ambient temperature.  Initially, some of the
increase in mean entropy comes from the removal of low-entropy gas through
condensation \citep{2001Natur.414..425V, 2002ApJ...576..601V,
2007ApJ...668....1N}.  However, the mean entropy at $\gtrsim 10$~kpc continues
to rise during the second Gyr of the simulation, after condensation of cold gas
has leveled off.  This rise is due to continual input of thermal energy by AGN
feedback, a small fraction of which escapes the inner few kpc and propagates into the
ICM, causing pressure-driven expansion of the ambient medium.  The right-hand
panel of Figure~\ref{fig:fkinetic_edot} shows that AGN feedback in the $f_{\rm
k} = 0$ simulation has injected $\sim 10^{64}$~erg after 2~Gyr, which is
comparable to the binding energy of the entire intracluster medium, although
most of this is radiated away by the cold gas. The fraction that does escape the central clump of cold gas
slows the condensation process by inflating the cluster core and driving the
cooling time of the ambient medium at $\sim 10$~kpc to $\sim 5$~Gyr but fails
to establish a self-regulated feedback loop.\par

\begin{figure}
   \includegraphics[width=0.49\textwidth]{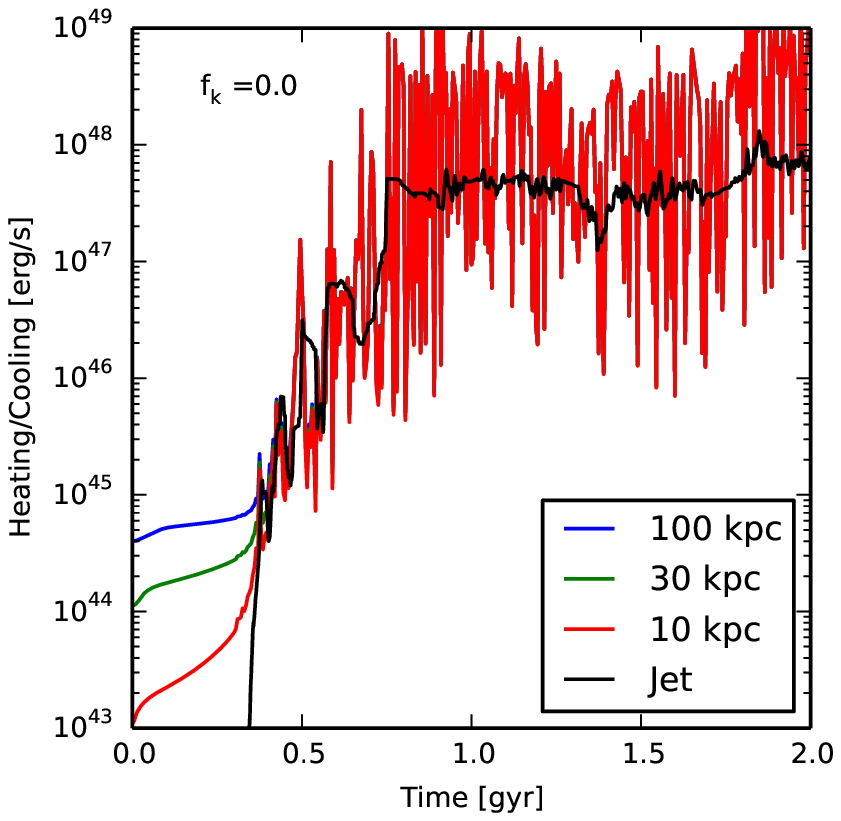}
   \centering
   \caption{The radiative cooling rate of the gas within different radii is
compared to the total jet power (thermal + kinetic) for the simulation with
$f_{\rm k}=0.0$. At times greater than 0.5 Gyr, all of the cooling is occuring within 10 kpc, and the three lines overlap.}
   \label{fig:thermal_cooling_balance}
\end{figure}

Despite the high AGN power, this simulation is not able to prevent a buildup of
cold gas for two reasons: (1) feedback energy does not propagate far enough
from the center, and (2) thermal feedback cannot destroy a large cold-gas
reservoir, once it develops. As illustrated in Figure
\ref{fig:thermal_cooling_balance}, almost all of the cooling comes from the
central 10 kpc where there is a concentration of cold gas with a very short
cooling time. Although gas at the temperature floor does not cool, a small rise
in temperature greatly increases its cooling rate and prevents the cold gas
from heating to the ambient temperature. \par

\begin{figure}
   \includegraphics[width=0.49\textwidth]{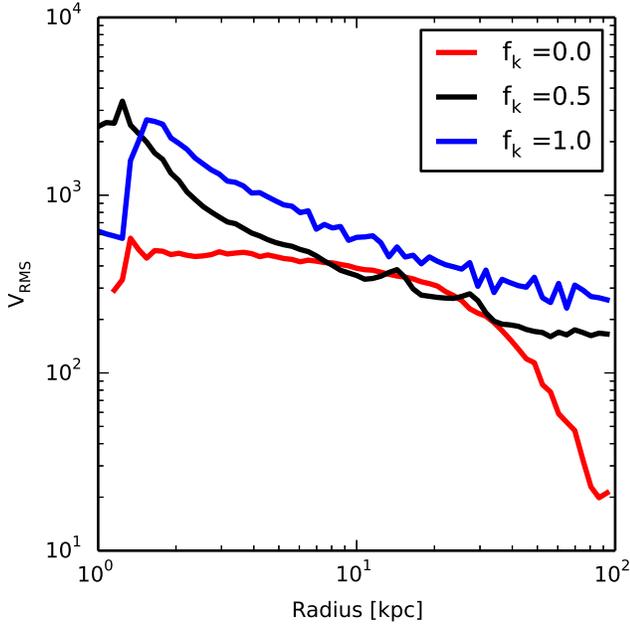}
   \centering
   \caption{ Mass weighted profile of RMS velocity in the hot ambient medium
for simulations with cold gas triggering and different kinetic fractions. All
profiles are computed 1.75 Gyr after the beginning of the simulation.}
   \label{fig:absvel_plot}
\end{figure}

The development of a large cold-gas reservoir is closely related to the failure
of thermal feedback to propagate feedback energy beyond the central $\sim
30$~kpc.  Figure \ref{fig:absvel_plot} shows the rms gas velocity as a function
of radius in simulations with different proportions of kinetic feedback
($f_{\rm k} =\, $0.0, 0.5, and 1.0, respectively).  In the pure thermal case,
there is a sharp drop in rms velocity beyond $\sim 30$~kpc which is not seen in
the simulations having some kinetic feedback.  Apparently, kinetic feedback is
more effective at transporting feedback energy to large radii.\par

This discrepancy arises because outward propagation of centrally injected
thermal feedback is limited by the amount of entropy it can generate.  It
creates central bubbles of hot gas which can buoyantly rise only until they
reach a layer of equivalent entropy.  Then the bubbles blend with their
surroundings.  In this set of simulations, centrally injected hot bubbles stop
rising and blend with the ambient medium at $\sim 30$~kpc, as indicated by the
rms velocity curves in Figure \ref{fig:absvel_plot}, as well as the propagation
of tracer fluid in panel H of Figure \ref{fig:profiles_vs_time}.  We therefore
conclude that our implementation of pure thermal feedback does not add much
heat to gas in the 30--100~kpc range of radii but instead steadily raises the
entropy of ambient gas at 10--30~kpc, which flattens its entropy gradient.
Kinetic feedback, on the other hand does propagate beyond 30~kpc and
consequently allows gas in the entire 10--100~kpc range to settle into a
quasi-steady, self-regulated state.\par

\subsubsection{Jet Precession}

\begin{figure}
   \includegraphics[width=0.49\textwidth]{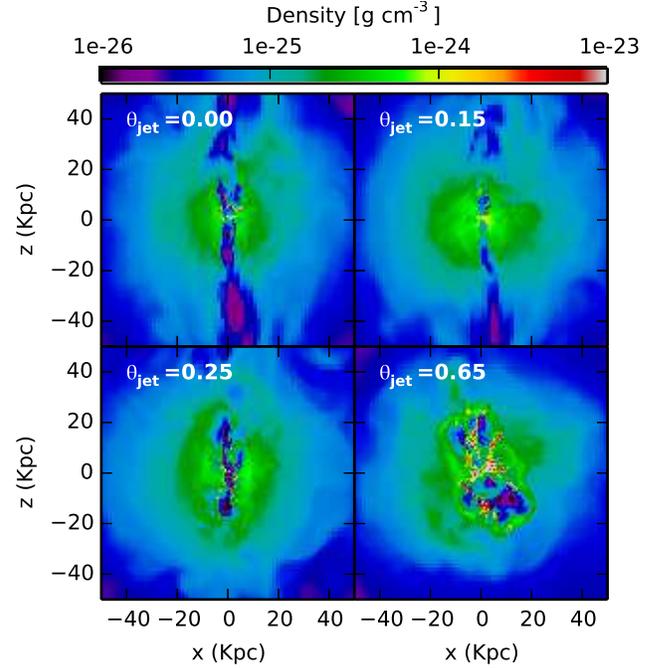}
   \centering
   \caption{Density slices for simulations with different values of
   $\theta_{jet}$. In our model, the AGN jet precesses around the $z$ axis with
   a period of 10 Myr at a constant angle $\theta_{jet}$ with the $z$ axis. All
   simulations use cold gas triggering and have $f_{\rm k}=0.5$.}
   \label{fig:precession_comparison}
\end{figure}

In addition to the breakdown between kinetic and thermal feedback, we have also
investigated the role of jet precession.  Jets from AGN can reorient themselves
on timescales of a few tens of Myr \citep{2006MNRAS.366..758D,
2013ApJ...768...11B}, but the details of this process are still uncertain, and
our subgrid model and idealized setup are not capable of self-consistently
modeling jet precession. Instead, we follow \citet{2014ApJ...789...54L} and
force the jets to precess around a fixed axis. Previous studies have found that
some precession is necessary for self-regulation if the jets are highly
collimated.  Otherwise, they drill long, narrow channels through the ICM and
deposit the bulk of their energy far from the zone in which self-regulation can
happen \citet{2006ApJ...645...83V}. \par

Figure \ref{fig:precession_comparison} shows slices of four simulations
performed with different jet precession angles. When the jets do not precess
($\theta_{jet}=0$), they carve channels through the ICM that extend well beyond
$\sim 40$~kpc.  However, due to Kelvin-Helmholz instabilities and artificial viscosity in
the ZEUS code, they still produce some heating close to the AGN but do not
drive strong shocks. With a small precession angle ($\theta_{jet} = 0.15,
0.25$), each jet continually encounters cold clouds of condensing material that
block its path. These jet-cloud interactions randomly divert the jets,
depositing their energy in a wider range of directions, which causes more of
their kinetic energy to thermalize at smaller radii. Precession also produces
more turbulence and creates shocks that propagate outward over a large range of
solid angles.  As the precession angle increases, the jet energy spreads over a
larger range of solid angles at ever smaller radii, and jet-cloud collisions
become more frequent. As seen in the last panel of Figure
\ref{fig:precession_comparison}, this leads to a more disturbed morphology at
$\lesssim 20$~kpc and a larger mass of accumulated cold gas. In that respect,
kinetic feedback with a very large precession angle becomes more like thermal
feedback, in that feedback energy does not propagate as far from the center
before it becomes thermalized.\par

\subsection{AGN Triggering Mechanisms}

\begin{figure*}
   \includegraphics[width=\textwidth]{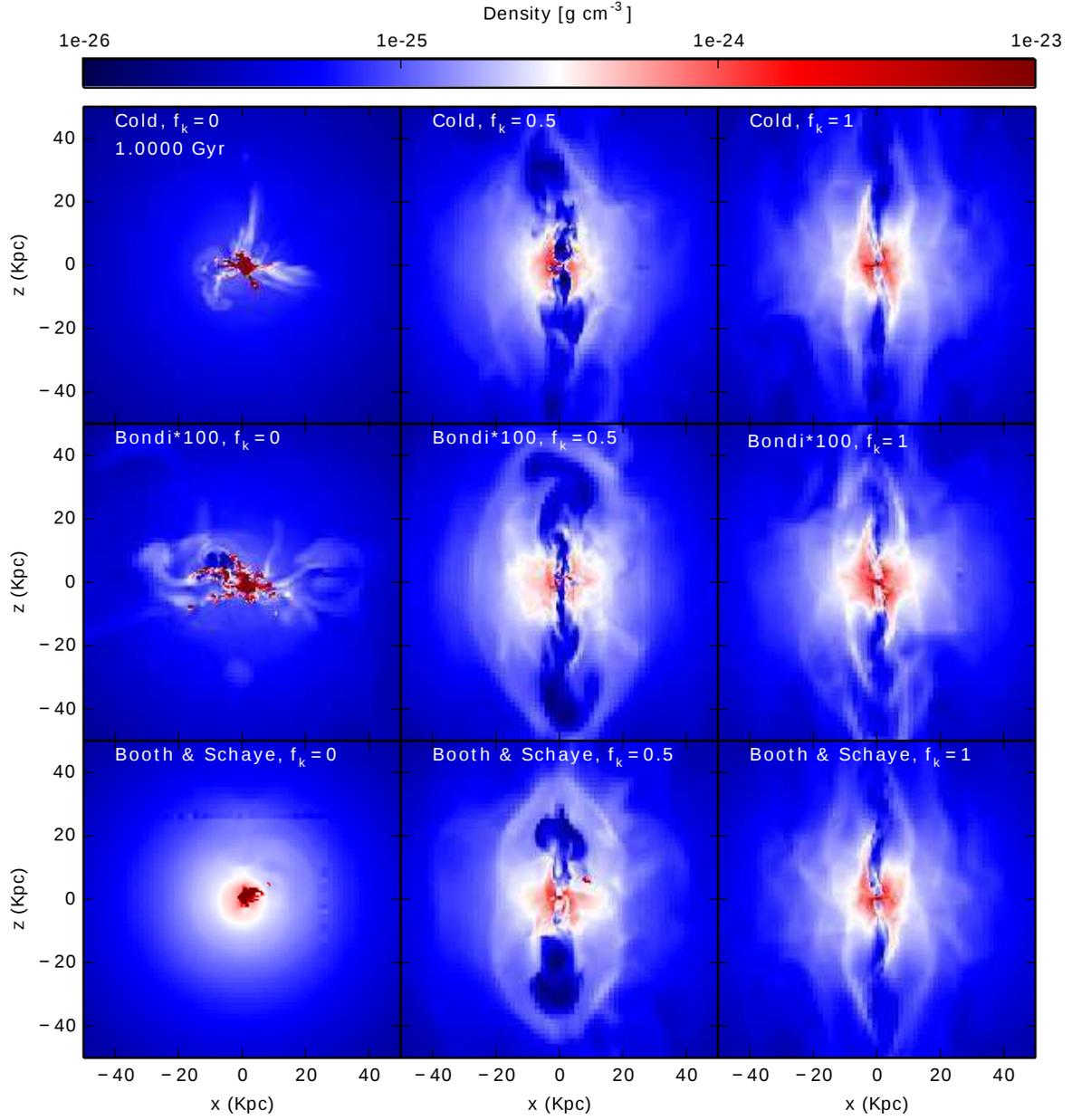}
   \centering
   \caption{Slices of density through 9 simulations with different triggering
   mechanisms and kinetic feedback levels. All simulations are shown 1 Gyr after
   the beginning of the run. Simulations in the top row are triggered by cold
   gas accretion, the middle row by Bondi accretion with a constant boost
   factor, and the bottom row using the method of \BS. $f_{\rm k}$ gives the fraction
   of the feedback that is returned as kinetic energy.}
   \label{fig:all_comparison}
\end{figure*}

\begin{figure*}
   \includegraphics[width=0.95\textwidth]{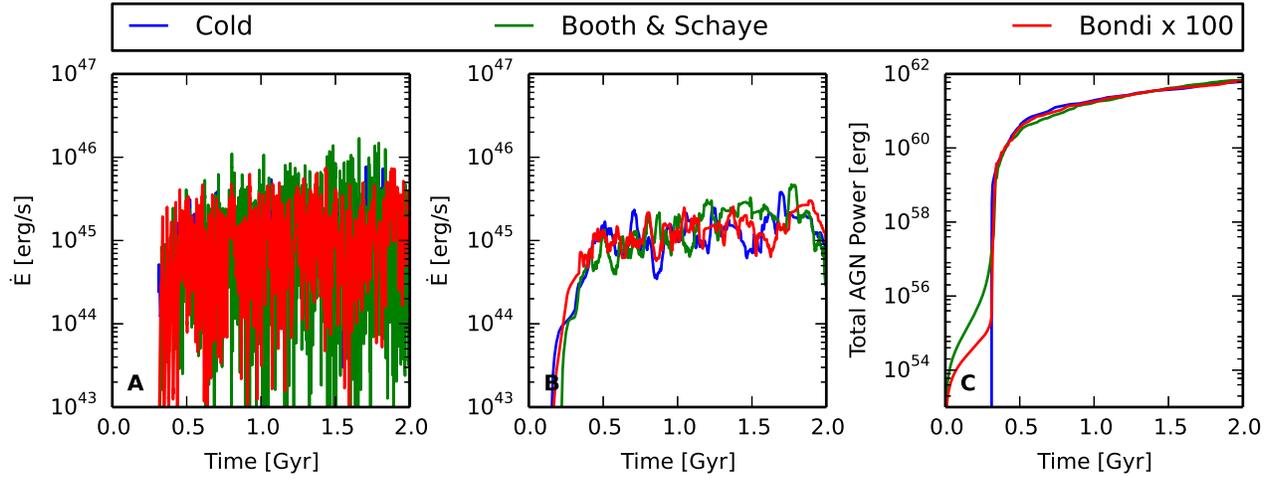}
   \centering
   \caption{Like Figure \ref{fig:fkinetic_edot}, but for simulations with
   different triggering mechanisms. All simulations have $f_{\rm k}=0.5$. As in
   Figure \ref{fig:fkinetic_edot}, Panel A shows the instantaneous value of
   $\dot{E}$, Panel B shows $\dot{E}$ smoothed over 50 Myr, and Panel C shows
   the cumulative jet power.}
   \label{fig:triggering_edot_comparison}
\end{figure*}

\begin{figure*}
   \includegraphics[width=\textwidth]{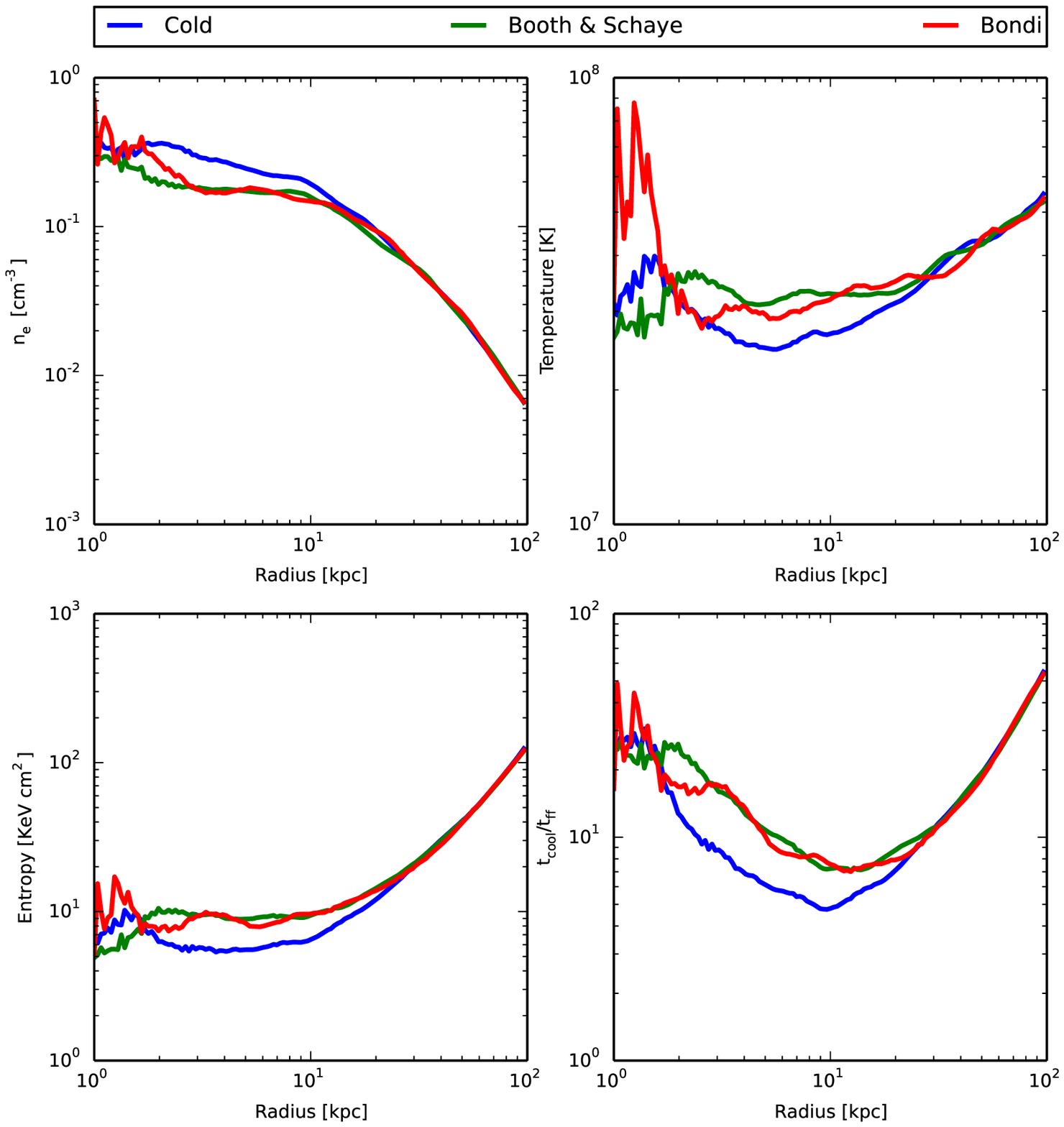}
   \centering
   \caption{Profiles of different quantities for simulations with different
   triggering mechanisms after 2 Gyr. All simulations use $f_{\rm k}=0.5$. $n_e$ is
   weighted by volume, while the other profiles are weighted by mass. All
   profiles excise gas below $3\times10^4$ K.}
   \label{fig:triggering_profiles}
\end{figure*}

We do not see strong differences between our simulations with different AGN
triggering mechanisms, as long as we are using a maximum spatial resolution of
196~pc (see Figure~\ref{fig:all_comparison}).  This is likely a consequence of
being able to resolve the multiphase medium in the region surrounding the AGN.
Since all of the triggering mechanisms considered are dependent on gas density,
a cold, dense clump of gas accreting will trigger a large outburst regardless of
the details of the triggering algorithm. The outburst will continue until the
cold gas is gone, ensuring that roughly the same amount of energy is released in
all cases. The cold gas would not necessary be resolved by simulations with
coarser resolution, implying that those simulations might be more sensitive to
the choice of triggering algorithm.\par

Each triggering mechanism depends on gas density either directly (BS and Boosted Bondi-like accretion) 
or indirectly (cold gas triggering), and in the BS and Boosted Bondi-like cases the boost
parameters have been chosen to provide the ``right'' amount of feedback. 
Due to the density-dependent accretion rate, a cold clump falling into the
accretion zone then always produces a surge in jet power that continues until
the clump is either completely heated or completely accreted.\par

Figure \ref{fig:triggering_edot_comparison} shows total jet power (thermal +
kinetic)  versus time for simulations with different triggering mechanisms.  In
the Bondi-like and BS runs, feedback is always active, but the power level is
relatively low before cold gas starts to condense and drives up the AGN
accretion rate.  After condensation begins, all three triggering mechanisms
lead to self-regulated jet power levels that are nearly identical.  Figure
\ref{fig:triggering_profiles} shows the radial profiles of various quantities
in the ambient hot ICM after 2~Gyr for each triggering mechanism.  There are
some differences in the inner 10 kpc, but this zone is strongly affected by the
quickly varying jet, producing profiles that are variable with time (see Figure
\ref{fig:profiles_vs_time}).  Between 10 and 30 kpc, the run with cold-gas
triggering is slightly colder and more susceptible to thermal instability than
the other runs, based on the lower $t_{\rm cool}/t_{\rm ff}$ ratio. Beyond 30
kpc, there are no significant differences between runs with different
triggering mechanisms.\par

\subsection{Accretion Radius}

\begin{figure*}
   \includegraphics[width=0.99\textwidth]{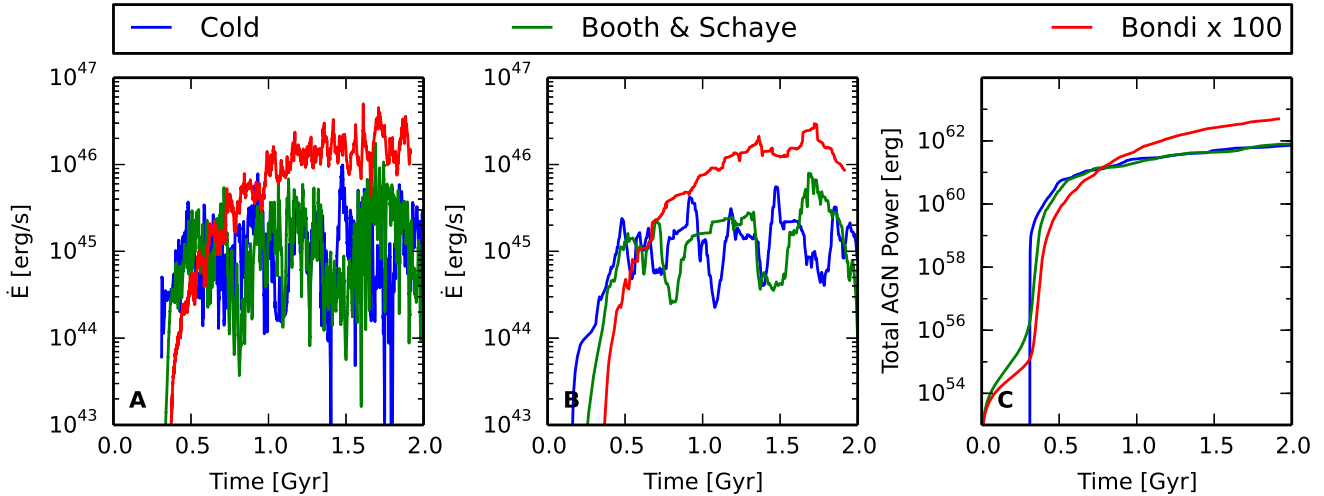}
   \centering
   \caption{Jet power vs. time  for simulations with
   $r_{acc}=r_{disk}=2$ kpc. All simulations have $f_{\rm k}=0.5$. Like in
Figure \ref{fig:fkinetic_edot}, Panel A shows the instantaneous value of
$\dot{E}$, Panel B shows $\dot{E}$ smoothed over 50 Myr, and Panel C shows the
cumulative jet power.}
   \label{fig:racc_edot}
\end{figure*}

In cosmological simulations of galaxy cluster evolution, one would like a
subgrid model for AGN feedback that gives reliable results for the SMBH
accretion rate even when the Bondi radius (let alone the Schwarzschild radius)
is not resolved.  At spatial resolutions coarser than $\sim 0.5$~kpc, the size
of the ``accretion zone'' that determines AGN feedback power will necessarily be
larger than that used in our fiducial simulations.  With this increase in
$R_{\rm acc}$, the responses of AGN triggering algorithms will depend on
conditions at larger radii, which can couple AGN feedback to ICM properties at
greater distances but may also permit larger amounts of gas to condense before
the AGN feedback response becomes strong enough to oppose cooling.

To understand how the size of the accretion zone affects AGN triggering, we
have carried out simulations in which $R_{\rm acc}$  is increased to 2 kpc.
The maximum spatial resolution remains the same, with a smallest cell width of
196 pc, meaning that the accretion radius is always resolved by multiple cells.
In carrying out these simulations we set the distance $R_{\rm J}$ of the
disk-shaped jet injection region equal to $R_{\rm acc}$, so that the jet
emanates from the edge of the accretion sphere, not from within it.\par

Figure \ref{fig:racc_edot} shows jet power as a function of time for the
simulations with a larger accretion radius. In all three simulations,
fluctuations in jet power are noticeably smaller than in the fiducial case.
With the exception of the Boosted Bondi-like run the cumulative jet power is
comparable to the earlier runs, and we did not observe quantitative differences
in the ICM properties.  However, the Boosted Bondi-like simulation, in which
AGN feedback power now depends on average gas properties within a larger
volume, takes longer to ramp up, resulting in a large ($>10^{12}$) mass of
cold gas and a higher cumulative jet power.\par

\section{Discussion}\label{section:discussion}

Our simulations have shown that different triggering and delivery methods for
subgrid models of AGN feedback can have profoundly different effects on the resulting
properties of the ICM. 
We are not attempting to determine which method is the most accurate model of an
AGN but rather to analyze the reasons for those differences. 
In each case, tracking the accumulation of cold-gas fuel is critical, meaning that 
we must consider what allows gas to transition from the hot ambient medium
into the cold-gas fuel reservoir.
In this section, we discuss that transition and consider the potential effects of 
physical processes not included in our models.\par

\subsection{Precipitation and AGN Fueling}

\begin{figure*}
   \includegraphics[width=0.99\textwidth]{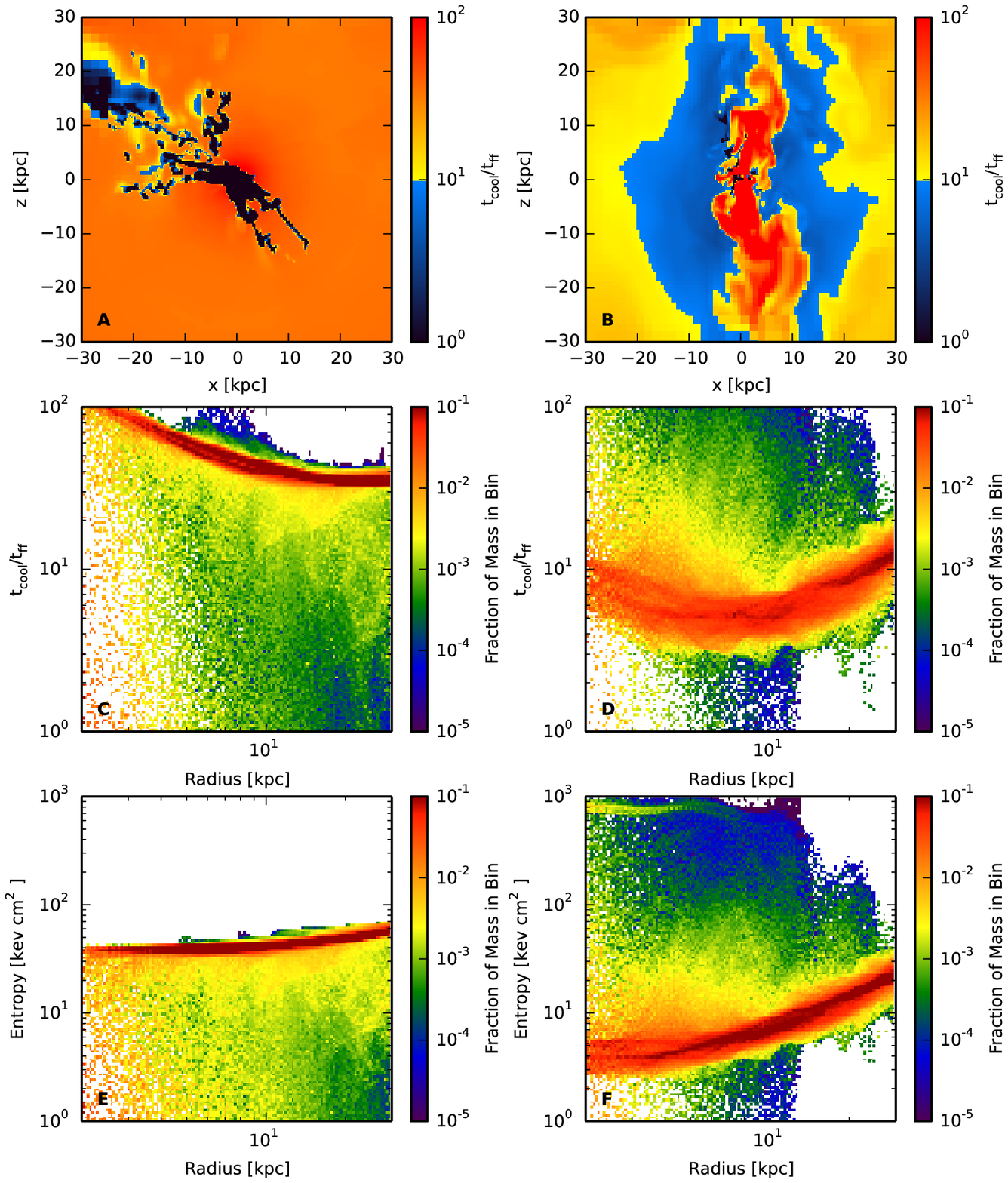}
   \centering
   \caption{Distribution of the $t_{cool}/t_{ff}$ ratio for simulations with
   cold triggering and either $f_{\rm k}=0.0$ (left column) or $f_{\rm k}=0.5$ (right
   column), shown at 1.46 Gyr after the beginning of the simulation.  Panels A
   and B show slices of the local $\TimeScaleRatio$ for each simulation.  The
   color-break in the scale at $\TimeScaleRatio=10$ indicates the precipitation
   threshold identified by earlier studies. Panels C and D show the distribution
   of $\TimeScaleRatio$ values normalized by the total mass at each radius.  The
   colors show the mass in each bin divided by the total mass in that radial
   shell.  Similarly, Panels E and F show the Entropy distribution.  Cold gas
   ($<3\times 10^4$ K) is excluded from the analysis.}
   \label{fig:timescale_phase}
\end{figure*}

Clearly, our simulations with pure thermal feedback behave markedly differently
than simulations with kinetic feedback, even when $f_{\rm k}$ is small. The
pure thermal feedback runs experience a large buildup of cold gas that essentially
smothers the AGN, causing it to fight back with increasingly powerful bursts.
The ICM in the vicinity of the AGN is subject to both radiative cooling and
heating from mixing, dissipation, and shocks.  Thus, analyzing the thermal
stability of the ICM may give insight into the accumulation of cold gas and
help to explain the differences that arise from among these feedback
algorithms.\par

\citet{2015Natur.519..203V} presented evidence for a ``precipitation
triggered'' model for coupling the AGN power to the cooling rate of the ICM. In
the precipitation model, the cooling ICM becomes thermally unstable, leading to
the condensation of cold gas.  This cold gas is then accreted by the AGN,
triggering feedback.  The feedback heats the ICM, restoring thermal stability
and reducing further accretion. As cosmological simulations typically lack the
resolution to model the condensation process itself, the thermal instability
criterion can be used to predict the amount of cold gas available for
accretion. For a gravitationally stratified medium, one would expect that
thermal stability would be related to two natural timescales --- the cooling
timescale $t_{\rm{cool}}$ and the dynamical timescale $t_{\rm{ff}}$. Simulations
\citep{2012MNRAS.419.3319M, 2012MNRAS.420.3174S} find that the formation of
cold gas from a thermally unstable medium can occur whenever $\TimeScaleRatio
\lesssim 10$ (But see \citet{2015ApJ...808...43M}, which finds that
condensation can occur for larger values in some circumstances.) Similarly, the
observations  of \citet{2015ApJ...799L...1V} and \citet{2008ApJ...683L.107C}
show that clusters with $\TimeScaleRatio \lesssim 10$ are likely to exhibit
multiphase gas, while clusters above that ratio do not. \par

Figure \ref{fig:timescale_phase} shows the distributions of $\TimeScaleRatio$
and specific entropy ($K$) for simulations with pure thermal ($f_{\rm k}=0.0$)
and part kinetic ($f_{\rm k}=0.5$) feedback.  Panel A of Figure
\ref{fig:timescale_phase} shows that the ICM in the thermal feedback simulation
is divided into two phases.  First, there is a hot phase with $\TimeScaleRatio
\gg 10$ that occupies the bulk of the volume outside of 10~kpc.  Second, there
is a large accumulation of cold gas that nearly smothers the AGN.  The cold gas
mass builds up quickly and then stops growing when the cooling time in the hot
ICM rises to several Gyr.  Outbursts of thermal feedback sporadically propel
streamers and blobs of cool gas radially outwards from the AGN.  These cold
streamers travel out several tens of kpc before turning around and raining back
down onto the core.  Panel C shows that there is a large spread in
$\TimeScaleRatio$ and $K$ in the 10--30~kpc range. Most of the gas at
intermediate values of $\TimeScaleRatio$ does not represent condensation in the
usual sense.  Instead, it is gas in the boundary layers of the streamers that
is either cooling onto them or being heated by interactions with the hot
ICM.\par

As Panels B, D, and F of Figure \ref{fig:timescale_phase} illustrate, the gas properties 
of the ICM for the simulation with $f_{\rm k} = 0.5$ are very different.
The ICM has much lower mean values of $\TimeScaleRatio$ and $K$ at each radius
out to 30~kpc. 
Panel B is typical of the state of the cluster after the jet has formed, with the volume
in which $\TimeScaleRatio \lesssim 10$ occupying a roughly spherical 
region of radius $\sim 20$~kpc, excluding a hot channel near the jet axis. 
Overall radiative losses are nearly balanced by gentle shock heating over 
several cooling times. 
However, at radii of $\sim 10$~kpc, where $\TimeScaleRatio$ reaches a minimum value
$\lesssim 10$, we observe relatively small amounts of condensing gas. 
Consistent with \citet{2014ApJ...789..153L}, this condensation occurs at the jet/ICM 
interface where the jet generates non-linear entropy fluctuations by uplifting 
low-entropy gas close to the AGN to greater heights, 
where it then condenses and falls back toward the center. 
These condensates are then accreted by the AGN, powering the jets
and maintaining thermal balance in the cluster.\par

The dramatic differences in the behavior of the cold gas and the jet in these
simulations has to do with how the AGN distributes energy to the surrounding
gas. In the pure thermal feedback case, the gas heated by the AGN at first
tends to follow the path of least resistance, bypassing the denser gas near the
core.  This leads to an accumulation of cold gas with a very short cooling
time, which is able to absorb and reradiate the AGN feedback at later times.
The AGN injects energy very close to the center of the cluster, where it is
immediately radiated away by cold gas. In the ($f_{\rm k}$ simulation, the
outflow creates a hot cocoon around itself that rapidly rises. This outflow
lifts the central gas outward, which helps disrupt the cooling flow and pulls
some low-entropy gas upward along with the jets. The kinetic outflow also
allows the feedback energy to penetrate to larger radii ($>10$ kpc) and heat
the ICM at greater radii. This helps to maintain the balance of heating and
cooling globally and prevents the ICM from dividing into a hot and a cold
phase. The jet is able to heat gas further out, through mixing, turbulent decay
and weak shocks, which prevents the cooling time of a large fraction of the ICM
from going below $10 t_{\rm ff}$. Thus, a cluster with a warmer, less dense
core will require less energy input to regulate than a cluster with a cold,
dense core.\par

In addition to depositing feedback further out, the kinetic outflows allow
cooling gas to mix with the hot gas in the jet. This increases the cooling time
of the ICM and strongly inhibits the formation of more cold gas. The cold or
cooling gas that is not accreted is soon swept up in the jet, where it is
disrupted or heated. The jet thus prevents the cold gas from smothering the AGN,
allowing the feedback to heat the ICM rather than quickly radiating away. This
explains why a weaker AGN is able to regulate the ICM in the kinetic jet case
than in the more powerful pure thermal feedback case.\par

\subsection{Caveats and Additional Physics}
\label{section:additional_physics}

In this study, both our setup and our implementation of AGN feedback have been
simplified in order to focus on the essential features of coupling between the
AGN and the ICM.  Of course, the situation in real clusters is more complicated
than our model. In addition to these simplifications, there are a number of
possibly relevant physical processes that we have not included in our model,
both to simplify the problem and to reduce the computational resources required.
These processes and their potential effects are discussed in this section.\par

\subsubsection{Conduction}
Our simulations do not include thermal conduction, either isotropic or along
magnetic fields. From a theoretical point of view \citep{2015Natur.519..203V,
2008ApJ...681L...5V}, while conduction may well be important for regulating the
thermal state of warm-core clusters, cool-core clusters lie below the $t_{\rm cool}$
profile at which conductive transport can balance radiative losses.
\citet{2013ApJ...778..152S} has simulated cool-core clusters with thermal
conduction but without AGN feedback, and concludes that thermal conduction is
not able to prevent the cooling catastrophe on its own and does not have a
large impact on global cluster properties. However, conduction could well be
important for the precipitation theory, as strong conduction could smooth out
the perturbations that evolve into non-linear overdensities.
\citet{2014MNRAS.439.2822W} have investigated the effects of conduction on
thermal stability and found that conduction would need to be quite strong to
prevent condensation.\par

\subsubsection{Magnetic Fields}
The intra-cluster medium is known to be weakly magnetized
\citep{2002ARA&A..40..319C}. Overall, the magnetic field is believed to be
tangled and dynamically unimportant. However, magnetic fields may affect heat
transport in the core by making conduction anisotropic, as the electrons that
mediate conduction will travel more easily along field lines than perpendicular
to them. The importance of anisotropic conduction will depend on the magnetic
field configuration, the development of plasma instabilities, and stirring of
the plasma by galaxy motions or AGN outflows. A tangled magnetic field would be
expected to supress conduction to roughly 1/3 of the Spitzer value. However, a
weakly magnetized, conducting ICM with a temperature gradient might be
susceptable to either the magnetothermal instability
\citep[MTI;][]{2000ApJ...534..420B, 2001ApJ...562..909B, 2008ApJ...673..758Q} or the
heat-flux-driven buoyancy instability \citep[HBI;][]{2008ApJ...673..758Q,
2009ApJ...703...96P}. In cool-core clusters, the HBI would align the magnetic
field perpendicular to an outward temperature gradient, limiting the inward heat
flux. However, simulations such as \citet{2011ApJ...740...81R} have found that
anisotropic thermal conduction is not strong enough to reorient the magnetic
fields, and \citet{2015arXiv151205796Y} find that stirring by the AGN would
overcome the HBI, leading to conduction with an effectiveness of $>0.2$ times
the Spitzer value.\par

While not dynamically important on large scales, magnetic fields may affect the
precipitation and AGN feedback processes. \citet{2014MNRAS.439.2822W} found that
anisotropic conduction will not prevent condensation unless the field is very
strong. Magnetic fields may be stronger and dynamically important close to the
AGN, where jet induced turbulence and field injection from the jet may amplify
the magnetic field \citep{2009MNRAS.399L..49D, 2012MNRAS.419.2293S,
2011ApJ...740...81R}. Along the AGN jets, magnetic draping is thought to play an
important role in preserving cavities and cold fronts against disruption from Kelvin-Helmholz
instabilities \citep{2007MNRAS.378..662R,2008ApJ...677..993D}. The preservation
of cavities would change the mode of heat transport in the cluster, because inflating
cavities and rising bubbles would be better able to
stir turbulence, transport hot gas to larger radii, and dredge up
cold gas in their wake.\par

\subsubsection{Star Formation}
BCGs in many cool core clusters are observed to be forming stars
\citep{2008ApJ...681.1035O, 2010ApJ...719.1619O, 2015arXiv151107884L,
  2015arXiv150806283M}, 
but stellar feedback alone can not prevent the cooling catastrophe in cool-core clusters
\citep[e.g.][]{2013ApJ...763...38S}. Although we do not include star formation in our
model, \citet{2015ApJ...811...73L} use a setup very similar to our fiducial
model to perform an extensive investigation of the role of star formation in
regulating AGN feedback. One expects the star formation rate
(SFR) of a BCG to be related to the amount of multiphase gas present.
\citet{2015ApJ...811...73L} do see a correlation between AGN feedback and
the SFR. In those simulations, stellar feedback is less effective than the AGN
at heating the ICM but more effective at consuming cold gas. 
If the AGN is in a low-power state, a central reservoir of cold gas builds up and 
boosts the AGN power on a $\sim 100$~Myr timescale. 
AGN feedback then heats the ICM and slows the rate of gas condensation.
However, the AGN remains powerful until star formation consumes the cold
gas in the central reservoir on a $\sim 2$~Gyr timescale.  
Without cold clouds to fuel it, the AGN feedback power subsides, and another
cycle soon begins as the ambient medium once again cools and becomes thermally unstable.
Thus, the primary effect of star formation is to regulate the cycling behavior 
of the AGN on Gyr timescales.\par

\subsection{Comparison With Similar Studies}\label{section:results_comparison}
As the importance of AGN feedback has gained greater appreciation in recent
years, several studies have been carried out to investigate the best way to
implement AGN feedback in simulations. It is difficult to do a comprehensive comparison
between our results and those of previous studies as those works have generally
sampled a limited fraction of the AGN feedback parameter space or assume vastly
different initial conditions than we do here.\par

The chief aim of this paper is to better understand which aspects of AGN
feedback implementations are most decisive in determining the qualitative
consequences of sub-grid models for AGN feedback. With this in mind, we discuss
the major differences between our implementation and some AGN implementations
used in related studies of AGN feedback. Where possible, we compare our results
to those obtained using these other algorithms. Note that in addition to the
major differences discussed here, there are many other small differences in the
details of how AGN feedback is implemented and in the choices of physical models
considered. As demonstrated by Section \ref{section:results}, the results of an
AGN feedback simulation may be sensitive to seemingly small differences in
implementation, and caution should be taken when comparing one set of results to
another.\par

\subsubsection{Li \& Bryan 2012-2015}
The cluster and AGN model employed in our paper are largely an extension of the
Li \& Bryan simulations of AGN feedback \citep{2012ApJ...747...26L,
2014ApJ...789...54L, 2014ApJ...789..153L, 2015ApJ...811...73L}, with only small
changes to the cluster and jet model (although we extend the range of triggering
and feedback parameters). Both our study and theirs use \texttt{Enzo}.\par

Given the similarities of our setups, it is not surprising that our simulations
give similar results. Our maximum spatial resolution is slightly coarser (196 pc
vs.~60 pc), but we obtain similar behavior for similar choices of feedback
parameters. Our findings indicate that the Li \& Bryan results should be
relatively insensitive to variations in the triggering mechanism, the amount of
AGN precession, and the details of the accretion process. Both studies find that
the behavior of the AGN is relatively insensitive to the kinetic fraction of the
outflow as long as the kinetic fraction is non-zero. Our study does find that
the mass of cold gas formed depends strongly on the AGN implementation, but does
not affect the long term behavior of the simulation. We generally see a mass of
cold gas that is an order of magnitude less than what Li \& Bryan found in their
fiducial model but obtain a similar mass when we use the same set of parameters.
This variability in the cold gas mass is consistent with the parameter variation
studies in \citet{2014ApJ...789...54L}.\par

\subsubsection{Gaspari et al. 2011}
\citet{2011MNRAS.411..349G} simulate AGN feedback using the \texttt{FLASH} code
\citep{2000ApJS..131..273F}. They model an idealized version of the cluster
Abell 1795 within a static and spherically symmetric gravitational potential
using a set of physical processes similar to those used here.  The minimum
resolution in their study is 2.7 kpc. AGN feedback is modeled as a purely
mechanical jet with either cold or hot (Bondi-like) triggering and different jet
efficiencies. They also consider both steady and intermittent jets. For
Bondi-like triggering, the accretion rate is calculated from the properties of
gas within 5 or 10 kpc.  \citet{2011MNRAS.415.1549G} uses a similar AGN model
but a gravitational potential appropriate for a galaxy group.\par

\citet{2011MNRAS.411..349G} finds that both a cold gas triggered and a Bondi
triggered AGN implementation are able to balance radiative cooling and preserve
a cool-core state. The most successful cold gas model (model A3 in that paper)
is significantly more bursty than our simulations, with a duty cycle of only
6\%, resulting in only around 50 outbursts each with power on the order of
$10^{48}$ erg/s. The total injected energy after 2 Gyr is on the order of
$10^{61}$ erg, consistent with our results. We attribute the observed difference
in outburst power and duty cycle to the choice of accretion radius, where we use
0.5 kpc and they use 10 kpc. As seen in Figure \ref{fig:racc_edot} in our paper,
increasing the size of the accretion radius results in a larger variation in AGN
power. This follows from more cold gas being able to fit inside the larger
accretion zone and from the difficulty of expelling cold gas from a larger
gravitational well.\par

In agreement with our results, \citet{2011MNRAS.411..349G} finds that Bondi
feedback with a large averaging zone (10 kpc in their simulations) is not able
to halt the cooling catastrophe. Their model with an averaging zone of 5 kpc is
able to balance cooling over a long period of time. Unlike the cold gas
triggered case, the Bondi implementation results in a low power (order $10^{44}$
erg/s) jet with little variation in intensity. In our simulations, the Bondi and
cold-triggered implementations act similarly when using an accretion
radius/averaging zone of 0.5 kpc. We ascribe this to the higher resolution of
our simulations, which are able to resolve the cold gas directly.

\subsubsection{Yang et al. 2012}
\citet{2012MNRAS.427.1614Y} examines the effect of different AGN subgrid models
on observable properties of simulated galaxy clusters. They model an idealized
cluster with virial mass $1.5 \times 10^{14} \Msun$ and a polytropic equation of
state, also using \texttt{FLASH}. The minimum resolution of these simulations is
1.0 kpc.  The physical processes considered are again similar to ours, while the
AGN feedback model is somewhat different, consisting of either large (tens of
kpc) thermal bubbles offset from the core or jets with widths of a few kpc. The
accretion rate was determined from the Bondi rate, with a constant boost factor
ranging from 1 to 100 in different runs.\par

Although \citet{2012MNRAS.427.1614Y} do consider jets with pure thermal feedback
(as well as thermal bubbles originating near the AGN), they do not see the same
smothering behavior that we do. In fact, the gas in their simulations does not
become very dense, rarely exceeding densities of $n_e=10^{-1}$~cm$^{-3}$. We attribute
these differences to the finer resolution of our simulations, which allow us to
resolve the condensation process and the formation of cold gas near the AGN.\par

\subsubsection{Dubois et al. 2012}
\citet{2012MNRAS.420.2662D} compare thermal and mechanical feedback in
cosmological simulations using the code \texttt{RAMSES}. The simulations
generally have a minimum resolution of 1.52 kpc, but some runs have higher
resolution. The AGN power is determined using the \BS\ method. Thermal energy is
released in a sphere of a a few cells near the AGN, while kinetic feedback is
released in a jet. Similar to \citet{2012MNRAS.427.1614Y}, they do not observe
AGN smothering, but again employ a coarser resolution than we use in our
simulations.\par

\section{Conclusions}\label{section:conclusions}

We have carried out a controlled comparison of several commonly used sub-grid
implementations of AGN feedback. Our model treats the AGN as a particle sitting
in the core of an idealized cool-core cluster. The AGN is triggered based on
local conditions (either the amount of cold gas or the Bondi rate, with either a
fixed or a density dependent boost) and returns energy to the ICM as either
centralized thermal blasts, a kinetic jet, or a mix of thermal and kinetic
energy. Our main conclusions are:

\begin{enumerate}

\item Purely thermal feedback produces very different results than feedback with
even a small kinetic component. In the pure thermal case, the AGN is initially
unable to inhibit cooling immediately outside of the core, leading to a buildup
of cold gas. This gas smothers the AGN and immediately radiates away the
feedback energy, even if the feedback zone itself is heated to a high
temperature. This also results in heating of the ICM outside of the core through a
combination of shock heating and preferential condensation of low entropy gas.
Adding a kinetic component allows the AGN to propagate energy outside of the
core and prevents smothering of the AGN.\par

\item When some fraction of the feedback is returned as a kinetic jet, the AGN
is able to prevent the large accumulation of cold gas that results from a
cooling catastrophe.  Instead, AGN feedback self-regulates the ICM in a
quasi-steady state with $\TimeScaleRatio \sim 10$ at $\lesssim 20$~kpc. The cluster
core is cooler overall than the case with pure thermal feedback, but contains
much less cold gas around the AGN.

\item We do observe large differences between cold-gas triggered feedback,
boosted Bondi-like triggering or Booth and Schaye accretion, as long
as the ``accretion zone'' used to determine the AGN fueling rate is sufficiently small
($\sim 200$~pc).  This is probably because all three methods, by design, end up
triggering strong AGN feedback when cold clouds begin to accumulate in the
accretion zone. 

\item Increasing the size of the accretion zone (to 2~kpc) reduces short-term
variation in jet power but does not significantly alter the total amount of AGN
feedback or the global ICM properties in the cold-gas triggered or Booth and
Schaye cases.  However, the boosted Bondi-like simulation does not achieve
self-regulation, because AGN feedback does not ramp up fast enough to prevent a
cooling catastrophe, resulting in a large central accumulation of cold gas.

\item Very large jet precession angles distribute the AGN feedback energy,
making simulations with significant kinetic output behave more like simulations
having pure thermal feedback.  This happens because the kinetic energy does not
escape to large radii and thermalizes closer to the AGN when it is spread over
too large a solid angle.

\end{enumerate}

Further improvements to sub-grid feedback models and a better understanding of
the AGN feedback process are necessary for the next generation of galaxy and
galaxy cluster simulations. On a theoretical level, much work is currently being
done on the link between thermal instability, cold gas formation, and its role
in triggering feedback. Ongoing observations with Chandra and XMM-Newton as well
as future observations with the Smart-X and Athena missions will give a better
understanding how the AGN feedback process operates in real clusters. Finally,
new implementations must be developed for capturing the connection between AGN
and their environments. While the simulations in this work have a maximum
resolution of $\sim 200$~pc, cosmological simulations and simulations with more
complicated physics generally have $\gtrsim $~kpc resolution due to
computational resource limits. In future work, we will aim at translating the
results from this project into a sub-grid implementation that can be used at
these coarser resolutions.

\acknowledgments
\section{Acknowledgments}
The authors would like to thank Gus Evrard, Mateusz Ruszkowski, Greg Bryan, and
Yuan Li for helpful discussions during the preparation of this paper. This work
was supported in part by Michigan State University through computational
resources provided by the Institute for Cyber-Enabled Research. BWO was
supported in part by the sabbatical visitor program at the Michigan Institute
for Research in Astrophysics (MIRA) at the University of Michigan in Ann Arbor,
and gratefully acknowledges their hospitality. This work was supported in part
by NASA through grants NNX12AC98G, NNX15AP39G, and Hubble Theory Grants
HST-AR-13261.01-A and HST-AR-14315.001-A. The simulations were run on the NASA
Pleiades supercomputer under allocation SMD-15-6514. The active particle
framework, upon which our AGN implementation is based, was developed by Nathan
Goldbaum at NCSA, and we thank him for his support. \texttt{Enzo} and
\texttt{yt} are developed by a large number of independent researchers from
numerous institutions around the world. Their commitment to open science has
helped make this work possible.

\bibliographystyle{apj}
\bibliography{apj-jour,ms}

\end{document}